\documentclass[11pt]{article}

\usepackage[margin=1.1in]{geometry}
\usepackage{amsmath, amssymb, amsthm}
\usepackage{booktabs}
\usepackage{graphicx}
\usepackage{subcaption}
\usepackage{xcolor}
\usepackage[round]{natbib}
\usepackage[colorlinks=true, linkcolor=blue!60!black, citecolor=blue!60!black,
            urlcolor=blue!60!black]{hyperref}

\graphicspath{{figures/}}

\newtheorem{definition}{Definition}

\newtheorem{hypothesis}{Hypothesis}

\DeclareMathOperator*{\argmax}{arg\,max}

\title{Auditing Algorithmic Collusion\\
       from Strategy Graphs}
\author{Nicolas Eschenbaum\thanks{Swiss Economics. \texttt{nicolas.eschenbaum@gmail.com}}
   \and Janusz M.\ Meylahn\thanks{Department of Applied Mathematics,
        University of Twente. \texttt{j.m.meylahn@utwente.nl}}}
\date{\today\\[0.5em] \textit{Preliminary draft --- comments welcome}}

\begin{document}

\maketitle

\begin{abstract}
\noindent
Detecting algorithmic collusion is challenging because regulators often have limited access to firms' algorithms, training data, and market information. We study an intermediate-information regime in which an auditor can query firms' frozen pricing policies and construct the induced strategy graph. Using a complete characterization of Nash equilibria in a repeated pricing game, we identify graph-theoretic features of strategy graphs that are associated with collusive reward-and-punishment schemes, including maximum betweenness, attractor in-degree, and average path length. We then test these metrics on policies learned by decentralized Q-learning and the Q-learning algorithm of \cite{calvano2020}. We find that especially the maximum betweenness and attractor in-degree are strongly correlated with the standard profit-based Collusion Index. Importantly, the proposed metrics rely only on the unlabeled topology of strategy graphs and require neither price histories, demand estimates, nor competitive and monopoly benchmarks. Our results suggest that the structure of frozen pricing policies contains robust signals of collusion among reinforcement learning algorithms and provides a promising basis for auditing algorithmic pricing systems under limited information.


\end{abstract}

\section{Introduction}\label{sec:intro}

Firms increasingly delegate pricing decisions to learning algorithms. A central concern is that such algorithms may sustain supra-competitive prices without an agreement or instruction to collude. Reinforcement-learning agents have been shown to do so in canonical oligopoly simulations \citep[e.g.,]{calvano2020, klein2021, asker2024}. Empirical work also associates algorithmic-pricing adoption with higher margins in German retail gasoline markets \citep{assad2024} and the use of common rental-pricing software, including RealPage products, with higher rents in U.S. multifamily markets during periods of economic recovery \citep{calderwang2026}. Together, these findings make detecting algorithmic collusion a practical concern. A central question in this regard is what information an authority needs to identify a collusive pricing policy.

At one extreme, a regulatory authority may have access to the entire code of the learning algorithms acting in a particular environment as well as full pricing histories and training data. At the other, the authority may only have access to (part of) the pricing history. The tools available for detecting algorithmic collusion will depend on the informational regime the authority is in. In the former, detecting algorithmic collusion is likely to be easier than in the latter. What information is available to the authority is itself a matter of regulation, reflecting a trade-off between the authority's ability to detect collusion and firms' interests in maintaining the confidentiality of proprietary data, algorithms, and business practices. In designing detection tools, we thus need to take this trade-off into account by identifying reliable statistical methods requiring minimal information. 

In the low-information regime, a price trace records one path through a learned policy, but not how the policy would respond from states that were not observed. Moreover, it is challenging to distinguish between price movements that reflect changes in the underlying market state (such as demand or remaining capacity) from strategic, potentially collusive patterns. Recent work attempts to develop statistical audits using data collected by pricing algorithms \citep{hartline2024} to tackle this challenge. In the high-information regime, the addition of source code and training data provides substantial information, but may be difficult to obtain and interpret, and thus infeasible in practice. In this paper, we study an intermediate information regime: an authority can query the frozen pricing policies -- the greedy action selected at each market state -- through disclosure requirements or a sandbox environment. We ask whether these policy responses provide sufficient structural signals of collusion without knowledge of demand, numerical price benchmarks, or training methodology and data.

The particular object potentially containing this structural signal is the \emph{strategy graph} induced by the frozen deterministic joint policy. Its nodes are market states, and each node has one edge to its successor under joint greedy play. Such graphs are called functional relations for which each weakly connected component contains one cycle, or attractor, and directed rooted trees leading to it. We hypothesize the existence of a structural signal in these graphs since collusive equilibria must make deviations unattractive and may route return paths through punishment states. In the equilibria we analyze, this logic appears, for example, as stronger bottlenecks, longer paths back to the attractor, and fewer states entering the attractor directly than under competitive outcomes. These observations can be captured using graph-theoretic concepts.

To identify potential graph-theoretic metrics for the detection of signals of collusion, we consider a baseline environment in which we have access to all Nash equilibria. This is a pricing duopoly with a logit demand model in which each player has three possible prices, leading to 101 possible Nash equilibria \citep{meylahn2025,meylahn2023does, meylahn2024can}. Based on these, we identify five graph metrics that correlate strongly with the collusiveness of the equilibria as measured by the Collusion Index (CI) \citep{calvano2020}.  These are the (1) maximum betweenness, (2) attractor in-degree, (3) average path length, (4) basin fraction, and (5) the number of attractors. We find that, on the equilibrium set, maximum betweenness increases with the CI, attractor in-degree decreases, average path length increases, basin fraction increases, and the number of attractors decreases. We additionally report the closeness centrality of the attractor, a sixth metric closely related to the average path length, to which we assign no separate hypothesis. These results provide the hypotheses that we then take to learned policies.

The equilibria used to identify potential graph-theoretic metrics are not the outcome of a learning algorithm. We first apply the metrics to policies learned by the canonical Q-learning algorithm of \citet{calvano2020}, training two agents in its fifteen-price environment at discount factors from $0.1$ to $0.95$ to produce 1,800 policy pairs that span competitive and supra-competitive outcomes. Maximum betweenness, attractor in-degree, and average path length (strongly) correlate with the profit-based Collusion Index at the predicted signs, with absolute pooled correlations between $0.57$ and $0.67$. The first two measures correlate increasingly with the Collusion Index as the discount factor rises.

We then subject the metrics to two robustness tests. The first replaces the learning algorithm with Decentralized Q-learning \citep{arslan2017}, which converges for weakly acyclic games, on a coarser grid of five prices per player. We continue to observe the same signs and correlations of similar magnitude of our measures with the Collusion Index. The second varies the outcome rather than the algorithm. Instead of varying the discount factor to obtain competitive learned policies, we rematch independently trained policies. These are overfit to their training rival \citep{eschenbaum2022}, so after rematching collusion breaks down and agents play competitive policies. We find that the maximum betweenness and attractor in-degree keep their signs and correlations. In both, the basin fraction and the number of attractors become informative when competition fragments the state space. Throughout, the graph metrics use no observed price traces, numerical price labels, demand estimates, or competitive and monopoly benchmarks.

\paragraph{Related literature.} We relate to three strands of literature. The first is work on algorithmic collusion. Learning algorithms sustain supra-competitive prices in a range of simulated oligopoly environments \citep{calvano2020, klein2021, asker2024}. Subsequent work studies the mechanisms behind these outcomes and whether learned behavior survives changes in exploration or in the deployment environment \citep{banchio2022, lambin2024, abada2023, eschenbaum2022}. We take the final joint policy as given. Rather than asking how frequently learning produces collusion or which learning mechanism is responsible, we ask how the structure of more and less collusive policies differs.

The second is work on detection, where our audit question is closest to \citet{hartline2024} and \citet{zhou2026}. \citet{hartline2024} define plausible algorithmic non-collusion and develop a statistical audit based on data collected by pricing algorithms. \citet{zhou2026} combine trace-level diagnostics with a frozen-policy graph audit, classify attractors by their price level and basin size, and study interventions that remove supra-competitive attractors. We also treat the frozen policy graph as the audit object, but use it differently. Our metrics depend only on unlabeled graph structure, rather than the price level of an attractor, and are derived from a complete analytically known equilibrium set before they are applied to learned policies.

The third is the analytical foundation, which builds on convergence results for Decentralized Q-learning \citep{arslan2017} and the equilibrium analysis of pricing games in \citet{meylahn2025, meylahn2023does}. Meylahn uses best-response graphs and stochastic basins of attraction to characterize which equilibria the learning algorithm selects. We instead study the state-transition graph induced by a fixed pair of pricing strategies. The complete equilibrium set provides the ground truth from which we derive the graph metrics tested in the larger Q-learning environments.

\paragraph{Overview.} Section~\ref{sec:setting} defines the strategy graph, Section~\ref{sec:theory} derives the metrics, Section~\ref{sec:calvano} applies them to learned policies, and Section~\ref{sec:robustness} reports the robustness tests. Section~\ref{sec:discussion} discusses implications for auditing, and Section~\ref{sec:conclusion} concludes.

\section{Setting}\label{sec:setting}

We consider $n$ agents engaged in repeated price competition. In each period $t \in \{1, 2, \ldots\}$, each agent $i \in \{1, \ldots, n\}$ selects a price $p_i(t)$ from a finite set $\mathcal{A}_i$. The history of play up to period $t$ is $H_t = \{(p_1(s), \ldots, p_n(s))\}_{s=1}^{t}$.

Agents may condition their pricing decisions on past play, but only through histories truncated at some length $l$, that is, the last $l$ price profiles
\begin{equation}
    H_{[t-l:t]} = \{(p_1(s), \ldots, p_n(s))\}_{s=t-l+1}^{t}
    \;\in\; \mathcal{H}_l := \Big(\prod_{i=1}^{n} \mathcal{A}_i\Big)^{l}.
\end{equation}
Each agent employs a state space $\mathcal{S}_i \in \mathcal{P}(\mathcal{H}_l)$, where $\mathcal{P}(\cdot)$ denotes the set of all partitions of its argument. The state space specifies the class of conditional strategies available to the agent. For example, an agent may condition on histories of length $l' \leq l$, since $\mathcal{H}_{l'} \in \mathcal{P}(\mathcal{H}_l)$, or only on its own past prices and not on those of its rivals (as in \cite{eschenbaum2022}). At any point in time, each agent is in a state $s_i(t) \in \mathcal{S}_i$, and the joint state is $s(t) = (s_1(t), \ldots, s_n(t)) \in \mathcal{S} := \prod_{i=1}^{n} \mathcal{S}_i$. Not every element of $\mathcal{S}$ can necessarily occur in play, because at each point in time there is a single realized history. Agents with a common state space, for instance, always agree on the state they are in. We write $\hat{\mathcal{S}}$ for the set of joint states that can actually occur.

\begin{definition}[Strategy graph]\label{def:strategy-graph} Agents employ strategies $\pi_i(a \mid s)$ with a unique greedy action $\argmax_a \pi_i(a \mid s)$ in each state. Given a joint strategy $\boldsymbol{\pi} = (\pi_1, \ldots, \pi_n)$, the \emph{strategy graph} $G(\boldsymbol{\pi})$ has node set $\hat{\mathcal{S}}$ and a directed edge from each joint state to the joint state reached when all agents play their greedy action.
\end{definition}

Strategy graphs are \emph{functional relations}: every node has out-degree exactly one and self-loops are allowed. Each weakly connected component (WCC) of a functional relation consists of a unique cycle -- the \emph{attractor} -- together with trees directed toward it. All metrics considered in this paper are computed on the largest WCC of the strategy graph, and we refer to its cycle as the attractor $A$.

The collusiveness of a joint policy is measured by the Collusion Index
\begin{equation}
    \Delta(\boldsymbol{\pi})
    := \frac{\bar{\pi} - \pi^{N}}{\pi^{M} - \pi^{N}},
\end{equation}
where $\pi^{N}$ and $\pi^{M}$ denote per-firm profits at the static Nash and monopoly prices, and $\bar{\pi}$ is the average per-firm profit realized under $\boldsymbol{\pi}$.\footnote{For the equilibria of Section~\ref{sec:theory} the attractor is always a symmetric price pair, so both firms earn $\bar{\pi}$ even when the strategy profile is asymmetric.} A value of zero corresponds to competitive play and a value of one to monopoly pricing. Computing $\Delta$ requires knowledge of the competitive and monopoly benchmarks, which are likely not available to an auditor. The metrics we will define  below do not require them.

\section{The Anatomy of Collusive Equilibria}\label{sec:theory}

We start with the complete, analytically known equilibrium set of a three-price game and derive graph-theoretic metrics of collusion from the equilibrium logic of reward-and-punishment strategies. The metrics are then applied to learned policies in Section~\ref{sec:calvano}, and Section~\ref{sec:robustness} tests them on a second learning algorithm and on rematched policy pairs.

Consider the duopoly game of \citet{calvano2020} with the price set reduced to three prices. Two symmetric firms face logit demand: given prices $(p_1, p_2)$, firm $i$ sells
\begin{equation}
    q_i = \frac{e^{(a_i - p_i)/\mu}}
               {\sum_{j=1}^{2} e^{(a_j - p_j)/\mu} + e^{a_0/\mu}}
\end{equation}
and earns $\pi_i = (p_i - c_i)\, q_i$ per period. We use the baseline parameterization of \citet{calvano2020} with $a_0 = 0$, $a_1 = a_2 = 2$, $\mu = 0.25$, and $c_1 = c_2 = 1$. Each firm chooses from three prices, $\mathcal{A}_i = \{p_L, p_M, p_H\}$, where $p_L$ is the static Nash price, $p_H$ the monopoly price, and $p_M$ their midpoint.\footnote{This corresponds to the price-grid construction of \citet{calvano2020} with three prices and $\xi = 0$.} Firms have a memory of one period and observe both their own and  their opponent's price. In the notation of Section~\ref{sec:setting}, $l = 1$ and $\mathcal{S}_i = \mathcal{H}_1$ for both firms. The joint state is last period's price pair, and $\hat{\mathcal{S}}$ consists of the nine possible price pairs.

This game serves as our ground truth for two reasons. First, its Nash equilibria are known completely. Following the enumeration approach of \citet{meylahn2025}, 101 distinct equilibrium strategy profiles arise as the discount factor varies: 29 symmetric and 72 asymmetric. Each equilibrium is a deterministic map from the nine states into price pairs, that is, a strategy graph on nine nodes. Second, learning in this environment has convergence guarantees. The DQ algorithm of \citet{arslan2017} converges to a Nash equilibrium in weakly acyclic games, and \citet{meylahn2023does} establishes weak acyclicity for the pricing games studied there and characterizes the likelihood with which each equilibrium is learned. The equilibrium set is therefore the complete set of policies that a provably convergent learner can arrive at in this game, free of learning noise.

Based on the dominant outcome -- that is, the attractor of the largest WCC -- the 101 equilibria fall into one of three classes: \emph{competitive} ($\Delta = 0$, both firms at $p_L$; 28 equilibria), \emph{partial collusion} ($\Delta = 0.70$, both at $p_M$; 48), and \emph{full collusion} ($\Delta = 1$, both at $p_H$; 25). The partial-collusion type lies at $\Delta = 0.70$ rather than $0.5$ because the index is defined over profits and logit profit is concave along the price diagonal. 

Within what we term as the competitive class with $\Delta = 0$, however, collusive strategies can still be present. A supra-competitive price must be protected by punishment, and one way to do so is to revert to $(p_L, p_L)$ permanently -- a grim trigger. Play then never returns to the collusive price, which forms an isolated component while the competitive sink governs the dominant basin, so as measured on the largest WCC the equilibrium is indistinguishable from competition. In the three-price setting, twenty-seven of the 28 competitive-class equilibria are of this kind. Our metrics, computed on the largest WCC, do not identify them and instead focus on collusion that includes a return to cooperation.

\begin{figure}[t]
    \centering
    \begin{subfigure}[t]{0.32\textwidth}
        \includegraphics[width=\linewidth]{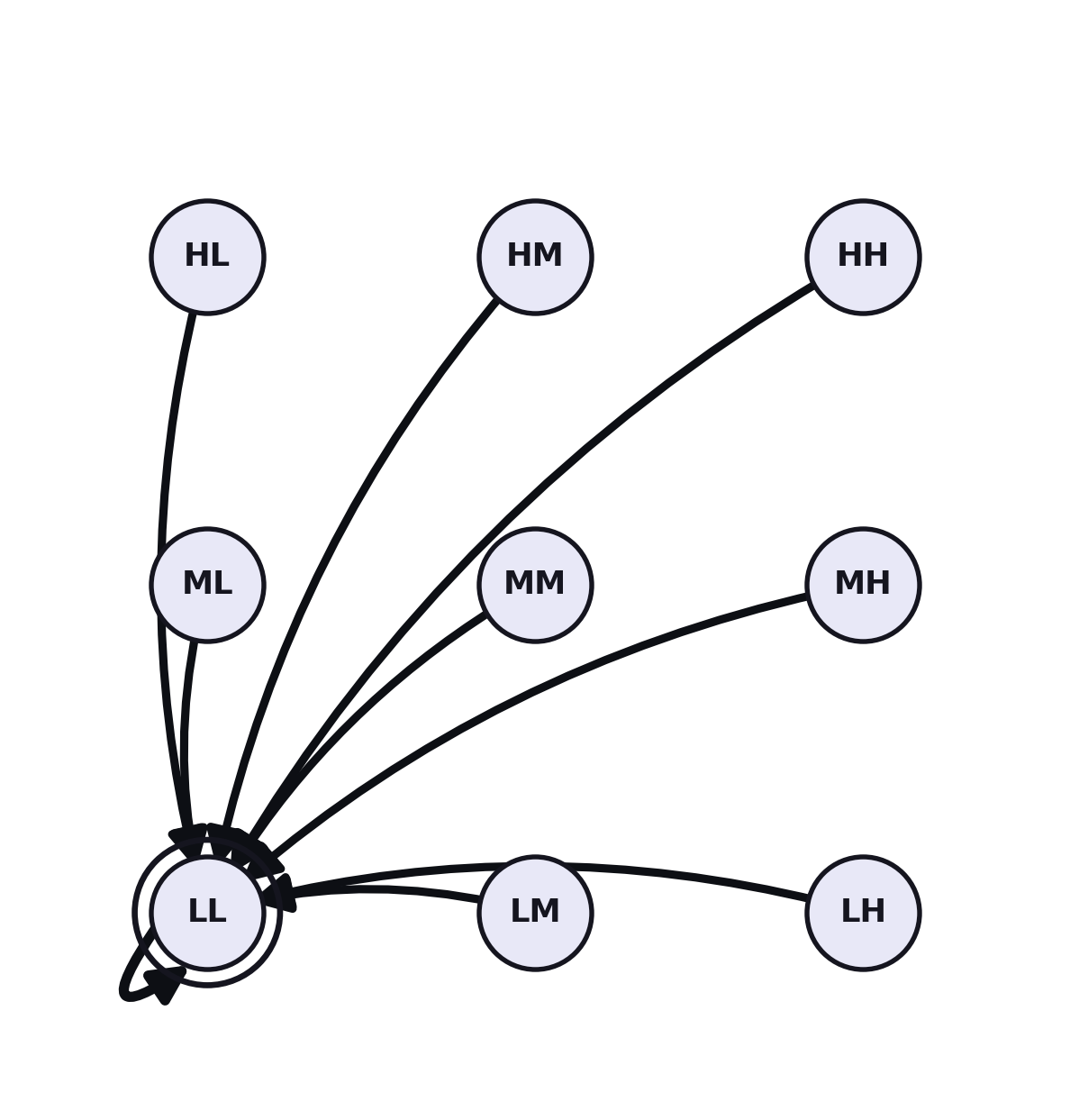}
        \caption{Competitive ($\Delta = 0$)}
    \end{subfigure}\hfill
    \begin{subfigure}[t]{0.32\textwidth}
        \includegraphics[width=\linewidth]{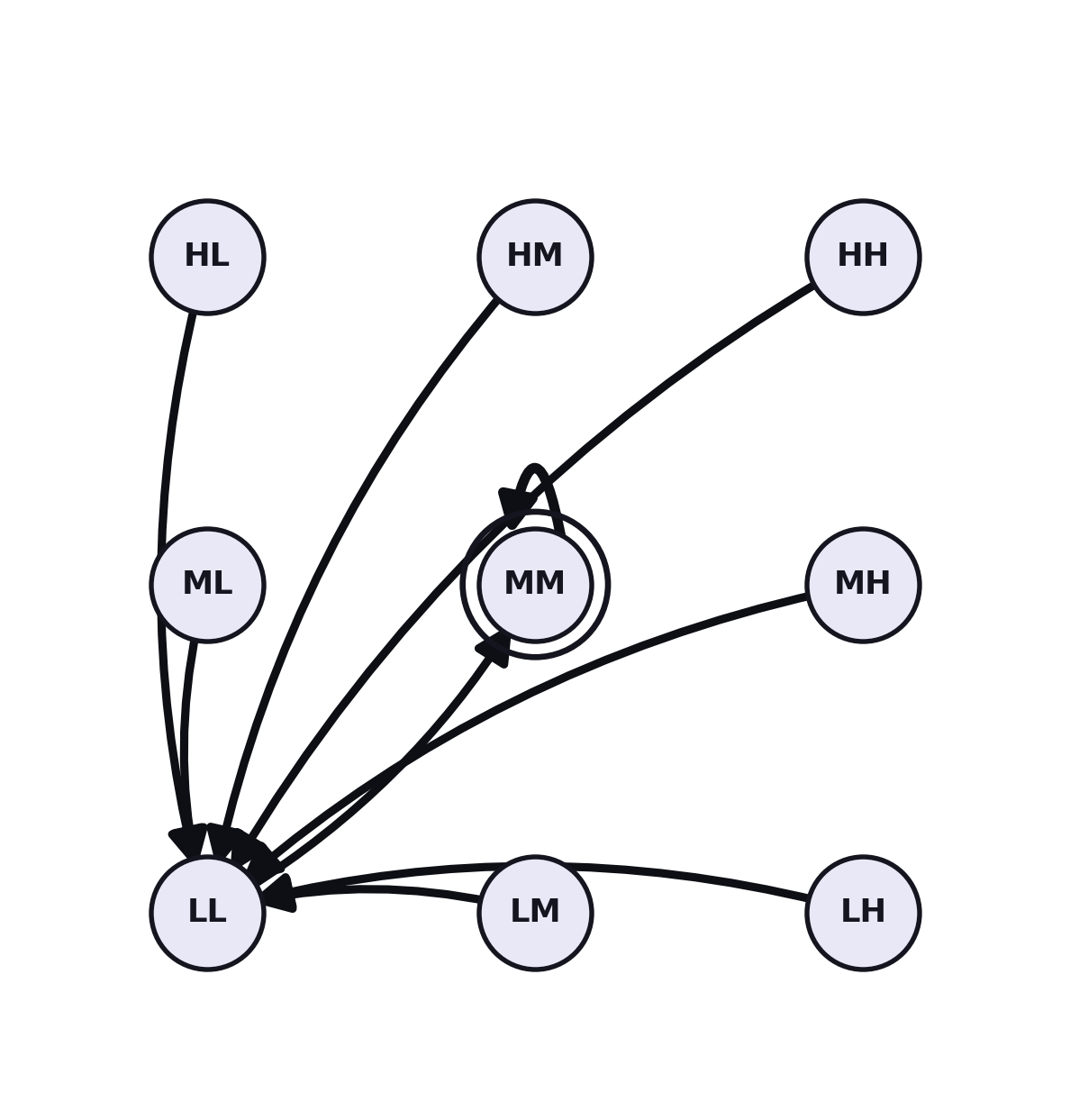}
        \caption{Partial collusion ($\Delta = 0.70$)}
    \end{subfigure}\hfill
    \begin{subfigure}[t]{0.32\textwidth}
        \includegraphics[width=\linewidth]{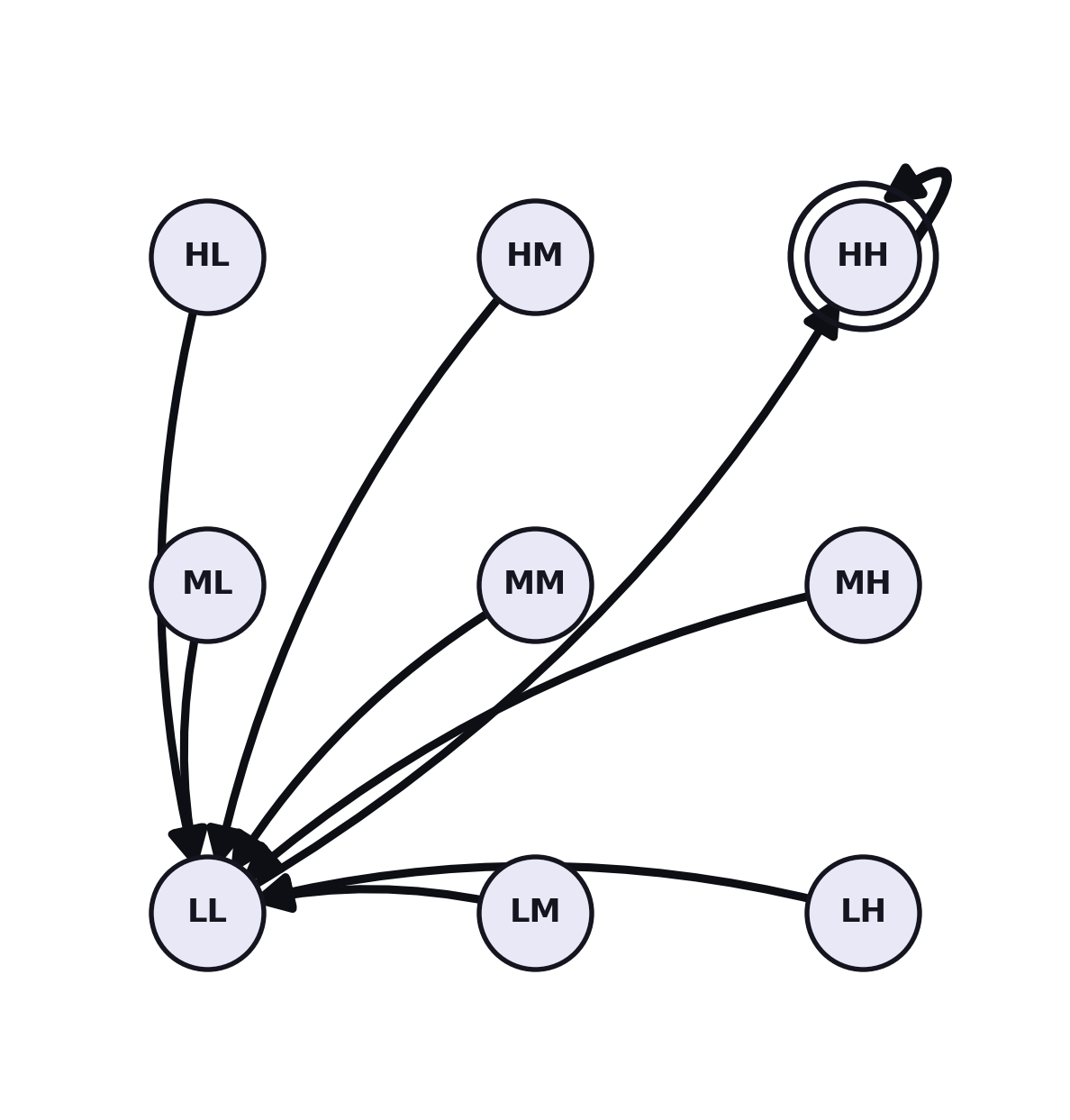}
        \caption{Full collusion ($\Delta = 1$)}
    \end{subfigure}
    \caption{Strategy graphs of three Nash equilibria of the three-price game. Nodes are joint states; arrows are the joint greedy transitions; the ringed node is the attractor of the largest WCC.} \label{fig:example-graphs}
\end{figure}

Figure~\ref{fig:example-graphs} shows the strategy graph of one equilibrium of each class. Panel~(a) shows a competitive equilibrium. Both firms price at $p_L$ in every state: pricing low is a best response to any history, so every state transitions directly into $(p_L, p_L)$, and play remains there. The attractor is a single state that collects the transitions of the entire state space in one step. No state lies on the path from any other state to the attractor, and any deviation from the attractor is corrected in one period under greedy play.

Panels~(b) and~(c) show collusive equilibria that sustain $(p_M, p_M)$ and $(p_H, p_H)$, respectively. A supra-competitive price can only be an equilibrium outcome if a unilateral deviation lowers the deviator's continuation payoff. The strategies must therefore respond to a deviation with low prices before the supra-competitive price is restored. In both panels, every state other than the attractor and the punishment state $(p_L, p_L)$ transitions into $(p_L, p_L)$, and from there play returns to the collusive attractor. Three differences to panel~(a) follow.

\begin{enumerate}
    \item The attractor is no longer reached from every state in one greedy step; apart from its own edge, only the punishment state transitions into it directly.
    \item Every other state reaches the attractor through the punishment state, which consequently lies on every return path in the graph.
    \item Returning to the attractor takes two steps rather than one.
\end{enumerate}

We formalize these differences using five graph-theoretic metrics. All metrics are computed on the largest WCC of the strategy graph. Let $V$ denote the node set of this component and let $A \subseteq V$ denote its attractor. Since the strategy graph is a functional relation, each node has a unique successor. We write $f:V \to V$ for this successor map, and for $u \in V$ we write 
\begin{equation}
\tau_A(u) := \min\{t \geq 0 : f^t(u) \in A\} 
\end{equation}
for the hitting time of the attractor. The unique path from $u$ to the attractor is therefore 
\begin{equation}
u, f(u), f^2(u), \ldots, f^{\tau_A(u)}(u). 
\end{equation} 
\begin{definition}[Maximum betweenness] 
For a node $v \in V$, define its return-path betweenness by 
\begin{equation}
b(v) := \left| \left\{ u \in V : v \in \{u, f(u), \ldots, f^{\tau_A(u)}(u)\} \right\} \right|. 
\end{equation} 
The maximum betweenness of the strategy graph is
\begin{equation}
B_{\max} := \max_{v \in V \setminus A} b(v). 
\end{equation} 
\end{definition} 

\begin{hypothesis}\label{hyp:betweenness}
The maximum betweenness $B_{\max}$is positively correlated with the Collusion Index.
\end{hypothesis}

\noindent \textbf{Intuition.} Collusive strategies must make deviations unattractive. A natural way to do so is to route many off-path states through a punishment state before returning to the collusive attractor. Such a punishment state lies on many return paths and therefore has high betweenness. Competitive strategies, by contrast, need not concentrate return paths through a single disciplining state: states can move directly to the competitive outcome. Hence, more collusive strategy graphs should exhibit stronger bottlenecks and higher maximum betweenness.  In panel~(a) no state lies on any path other than its own, so the maximum is $1$; in panels~(b) and~(c) every path runs through the punishment state, so the maximum is $8$.

\begin{definition}[Attractor in-degree] 
For $a \in V$, let 
\begin{equation}
d(a) := \left|\{u \in V : f(u)=a\}\right| 
\end{equation} 
denote its in-degree within the largest weakly connected component. The attractor in-degree is the average in-degree of the attractor nodes, 
\begin{equation}
D_A := \frac{1}{|A|} \sum_{a \in A} d(a). 
\end{equation} 
\end{definition} 

For comparability across strategy graphs of different sizes, we also report the normalized maximum betweenness $\tilde{B}_{\max} := B_{\max}/|V \setminus A| \in (0,1]$, the largest share of transient states that funnel through a single state. It equals one when one state lies on the return path of every transient node, and carries the same sign as Hypothesis~\ref{hyp:betweenness}.

\begin{hypothesis}\label{hyp:indegree}
The in-degree of the attractor $D_A$ is negatively correlated with the Collusion
Index.
\end{hypothesis}

\noindent \textbf{Intuition.} In a competitive strategy graph, many states may transition directly into the competitive attractor because immediate competitive pricing is already incentive compatible. In a collusive strategy graph, however, deviations often need to be followed by punishment before the collusive outcome is restored. As a result, fewer states enter the collusive attractor directly. The more collusive the outcome, the more important such detours become, and the lower the attractor in-degree should be. In panel~(a) all nine states enter the attractor directly; in panels~(b) and~(c) only the punishment state and the attractor itself do.

\begin{definition}[Average path length] The average path length to the attractor is 
\begin{equation}
L_A := \frac{1}{|V|} \sum_{u \in V} \tau_A(u). 
\end{equation}
Nodes on the attractor satisfy $\tau_A(u)=0$. 
\end{definition} 

\begin{hypothesis}\label{hyp:path}
The average path length $L_A$ is positively correlated with the Collusion Index.
\end{hypothesis}

\noindent \textbf{Intuition.} If play is competitive, deviations or off-path states can typically be corrected immediately, so most states are close to the attractor. In a collusive equilibrium, returning to the collusive outcome often requires an intermediate punishment phase. This increases the number of transitions needed to reach the attractor from off-path states. Therefore, more collusive strategy graphs should have longer average return paths. In panel~(a) every state is at most one step from the attractor; in panels~(b) and~(c) deviations return in two.

\begin{definition}[Basin fraction] 
The basin fraction is $\beta := \frac{|V|}{|\hat{\mathcal{S}}|}.$
\end{definition} 

\begin{hypothesis}\label{hyp:basin}
The basin fraction $\beta$ is positively correlated with the Collusion Index.
\end{hypothesis}

\noindent \textbf{Intuition.} A collusive strategy should restore the collusive outcome from a large portion of the state space. If some region of the graph were trapped in a different component, play could fail to return to the collusive attractor after certain deviations or histories. Collusive strategies therefore tend to organize the state space around a dominant basin of attraction. Competitive strategies do not require such global coordination, so the state space may be more fragmented across several components. 

\begin{definition}[Number of attractors] 
Let $\mathcal{C}(G)$ denote the set of weakly connected components of the strategy graph $G$. Since $G$ is a functional relation, each component contains exactly one directed cycle, or attractor. The number of attractors is therefore $K := |\mathcal{C}(G)|$.  
\end{definition} 

\begin{hypothesis}\label{hyp:wcc}
The number of attractors $K$ is negatively correlated with the Collusion Index.
\end{hypothesis}

\noindent \textbf{Intuition.} The reasoning mirrors that for the basin fraction. Sustained collusion requires play to spend most of its time in the collusive attractor and to return there after deviations. This is easier when the strategy graph is organized around a single dominant component. Multiple attractors indicate that different parts of the state space lead to different long-run outcomes, which is less consistent with a robust collusive scheme. Hence, more collusive strategy graphs should have fewer attractors.

We additionally report the closeness centrality of the attractor, the reciprocal of the average number of steps from the graph's states to the attractor. It is closely related to the path length and the in-degree, and we do not state a separate hypothesis for it. Note also that all three panels of Figure~\ref{fig:example-graphs} consist of a single component absorbing all nine states: the competitive equilibrium in panel~(a) satisfies the properties behind Hypotheses~\ref{hyp:basin} and~\ref{hyp:wcc} as well. These two metrics can only separate the classes through competitive equilibria that fragment into several components, a point we return to below.

Table~\ref{tab:anatomy} reports the metrics by equilibrium class in the three-price game. The results are in line with the hypotheses. The maximum betweenness equals $1$ in every competitive equilibrium and is at least $5$, of a possible $8$, in every fully collusive one. The average path length is at most $0.89$ in every competitive equilibrium and at least $1.33$ in every fully collusive one. The mean in-degree of the attractor falls from $6.75$ in the competitive class to $2.48$ under full collusion, and partial-collusion equilibria lie between the two classes on every metric.

\begin{table}[t]
\centering
\small
\begin{tabular}{lccc}
\toprule
Metric & Competitive ($\Delta = 0$) & Partial ($\Delta = 0.70$) & Full ($\Delta = 1$) \\
\midrule
Attractor in-degree & $6.75$\,{\scriptsize$[5,\,9]$} & $4.33$\,{\scriptsize$[2,\,7]$} & $2.48$\,{\scriptsize$[2,\,5]$} \\
Max.\ betweenness $\max_{v \notin A} b(v)$ & $1.00$\,{\scriptsize$[1,\,1]$} & $5.33$\,{\scriptsize$[3,\,8]$} & $7.52$\,{\scriptsize$[5,\,8]$} \\
Average path length & $0.85$\,{\scriptsize$[0.8,\,0.89]$} & $1.38$\,{\scriptsize$[1.1,\,1.7]$} & $1.96$\,{\scriptsize$[1.3,\,2.3]$} \\
Basin fraction & $0.75$\,{\scriptsize$[0.56,\,1]$} & $0.96$\,{\scriptsize$[0.89,\,1]$} & $1.00$\,{\scriptsize$[1,\,1]$} \\
Number of attractors & $2.21$\,{\scriptsize$[1,\,3]$} & $1.33$\,{\scriptsize$[1,\,2]$} & $1.00$\,{\scriptsize$[1,\,1]$} \\
Closeness centrality & $0.72$\,{\scriptsize$[0.5,\,1]$} & $0.62$\,{\scriptsize$[0.47,\,0.8]$} & $0.47$\,{\scriptsize$[0.38,\,0.67]$} \\
\midrule
$N$ & 28 & 48 & 25 \\
\bottomrule
\end{tabular}
\caption{Structural anatomy of the 101 Nash equilibria of the three-price game, grouped by the collusiveness of the dominant outcome. Cells report the mean and $[\min,\max]$ of each metric over the class. Metrics computed on the largest WCC.}
\label{tab:anatomy}
\end{table}

As anticipated, the basin fraction and the number of attractors are the least informative: every fully collusive equilibrium consists of a single component absorbing all nine states, but so does the competitive equilibrium of panel~(a), and the class ranges in Table~\ref{tab:anatomy} overlap. The correlations between the metrics and the Collusion Index across all 101 equilibria carry the hypothesized signs. Normalizing the maximum betweenness leaves the association essentially unchanged ($\tilde{B}_{\max}$: $\rho = +0.66$). We report them next to the learned-policy results in Section~\ref{sec:calvano} and, separately for symmetric and asymmetric equilibria, in Table~\ref{tab:theory} of Appendix~\ref{app:additionalTables}.

\section{Application to Learned Policies}\label{sec:calvano}

The hypotheses were derived from exact equilibria, but learned policies are not necessarily exact equilibria. We now consider the canonical Q-learning algorithm of \citet{calvano2020} to study whether the graph structure of Section~\ref{sec:theory} appears in learned policies and allows us to identify collusive policy pairs.

We consider the environment of Section~\ref{sec:theory}               with the price set enlarged to fifteen equally spaced prices that contain $p^N$ and $p^M$ exactly.\footnote{The grid of \citet{calvano2020} extends the interval $[p^N, p^M]$ by a factor $\xi = 0.1$ on each side and does not contain the two benchmark prices exactly. We set the step size to $(p^M - p^N)/12$, with one step below $p^N$ and one above $p^M$, so that both lie on the grid. Our results do not depend on the specific grid choice.}  Two Q-learning agents play the game for $T = 10^6$ periods per run. As in \citet{calvano2020}, each agent observes last period's price pair, chooses $\varepsilon$-greedily with $\varepsilon_t = e^{-\beta t}$ and $\beta = 4 \times 10^{-6}$, and updates its Q-matrix with learning rate $\alpha = 0.15$; Q-matrices are initialized at the discounted payoff from uniform random play. The joint state space $\hat{\mathcal{S}}$ consists of the $15^2 = 225$ price pairs.

We vary the discount factor $\gamma \in \{0.1, 0.3, 0.5, 0.7, 0.9, 0.95\}$ and simulate 500 independent runs for $\gamma = 0.1$ and $\gamma = 0.95$ and 200 for each intermediate value, 1800 runs in total. We do so because at $\gamma = 0.95$, the value studied by \citet{calvano2020}, learners consistently converge to supra-competitive outcomes: no run ends at competitive profit levels -- the minimum index across the 500 runs is $0.13$ -- so a simple replication would contain no competitive strategy graphs. A discount factor of $\gamma = 0.1$ in turn results in consistent (near-)Nash play and competitive strategy graphs. The mean Collusion Index rises across the range of discount factors, from $0.20$ at $\gamma = 0.1$ to $0.65$ at $\gamma = 0.95$.

For each run we compute the Collusion Index from the average profits over the final 100 periods, and the metrics from the strategy graph induced by the greedy actions of the two final Q-matrices.\footnote{The attractor reached by greedy play from the final state coincides with the largest WCC in 97.7 percent of runs.} Table~\ref{tab:gamma-theory} reports the correlations between the metrics and the Collusion Index for each discount factor, pooled across all runs, and, for comparison, on the equilibrium set of the three-price game.

\begin{table}[h]
\centering
\scriptsize
\begin{tabular}{lcccccccc}
\toprule
 & \multicolumn{8}{c}{$\rho(\cdot, \Delta)$} \\
\cmidrule(lr){2-9}
Metric & $\gamma{=}0.1$ & $\gamma{=}0.3$ & $\gamma{=}0.5$ & $\gamma{=}0.7$ & $\gamma{=}0.9$ & $\gamma{=}0.95$ & Pooled & \textbf{Equilibria} \\
\midrule
Max.\ betweenness centrality & $+0.064$ & $+0.142$ & $+0.268$ & $+0.484$ & $+0.702$ & $+0.733$ & $+0.669$ & $+0.941$ \\
Max.\ betweenness (norm.) & $+0.035$ & $+0.140$ & $+0.277$ & $+0.469$ & $+0.664$ & $+0.681$ & $+0.662$ & $+0.931$ \\
In-degree centrality & $+0.053$ & $-0.171$ & $-0.324$ & $-0.346$ & $-0.541$ & $-0.555$ & $-0.666$ & $-0.829$ \\
Number of attractors & $-0.144$ & $-0.044$ & $-0.052$ & $-0.219$ & $-0.463$ & $-0.509$ & $-0.131$ & $-0.746$ \\
Basin fraction & $+0.121$ & $+0.107$ & $+0.015$ & $+0.191$ & $+0.405$ & $+0.410$ & $+0.135$ & $+0.833$ \\
Closeness centrality & $-0.118$ & $-0.058$ & $-0.192$ & $-0.168$ & $+0.209$ & $+0.391$ & $-0.334$ & $-0.659$ \\
Average path length & $-0.035$ & $-0.044$ & $+0.147$ & $+0.352$ & $+0.407$ & $+0.239$ & $+0.566$ & $+0.869$ \\
\midrule
$\bar\Delta$ & $0.200$ & $0.203$ & $0.230$ & $0.337$ & $0.570$ & $0.650$ & $0.385$ & $0.580$ \\
$N$ & 500 & 200 & 200 & 200 & 200 & 500 & 1800 & 101 \\
\bottomrule
\end{tabular}
\caption{Pearson correlations $\rho(\cdot,\Delta)$ between strategy graph metrics and the Collusion Index: Calvano Q-learning by discount factor $\gamma$ and pooled, next to the analytically known equilibria of the three-price game (\emph{Equilibria}).}
\label{tab:gamma-theory}
\end{table}

Pooled across the 1800 runs, the maximum betweenness ($\rho = +0.67$), the attractor in-degree ($-0.67$), and the average path length ($+0.57$) correlate strongly with the Collusion Index and carry the hypothesized signs. The basin fraction ($+0.14$) and the number of attractors ($-0.13$) also carry the hypothesized signs but are weak, in line with the equilibrium analysis, where these two metrics were the least informative. The closeness centrality is negatively correlated in the pooled sample but not consistently so across discount factors.

The per-$\gamma$ columns show where the signal comes from. At $\gamma = 0.1$ hardly any run sustains supra-competitive profits, and no metric correlates with the index. As $\gamma$ rises and collusive outcomes appear, the correlations of the maximum betweenness and the attractor in-degree grow in magnitude, reaching $+0.73$ and $-0.56$ at $\gamma = 0.95$.

\begin{figure}[h]
    \centering
    \includegraphics[width=\textwidth]{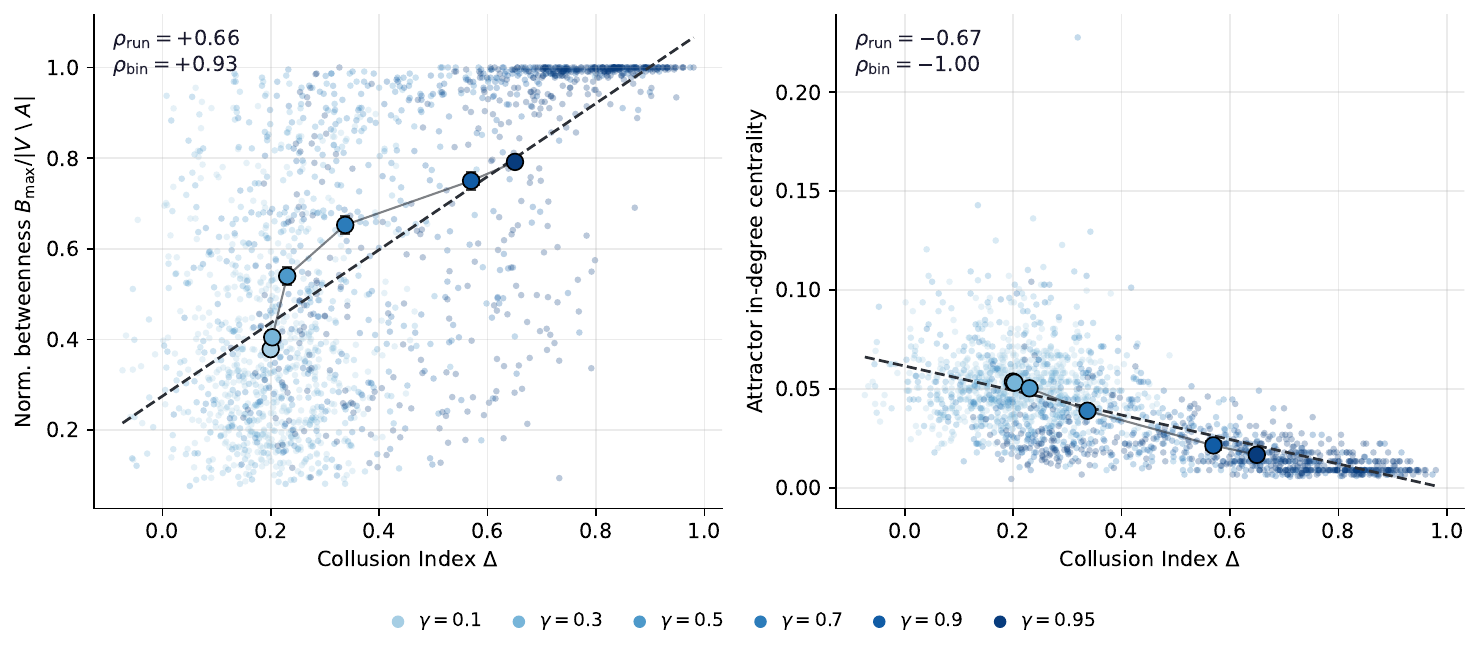}
    \caption{Collusion Index against the normalized maximum betweenness
    $B_{\max}/|V\setminus A|$ (left) and the attractor in-degree (right) across
    all 1800 Calvano Q-learning runs, shaded by discount factor $\gamma$.
    Dashed lines are pooled least-squares fits; large markers are the
    per-$\gamma$ means.}
    \label{fig:scatter}
\end{figure}

Figure~\ref{fig:scatter} plots the two strongest metrics against the Collusion Index across all runs, using the normalized maximum betweenness in the left panel. Runs with a high index concentrate near the top of the left panel and at low attractor in-degree. This concentration is not an artifact of the scale: among the $\gamma = 0.95$ runs with an index above $0.8$, the median maximum betweenness is $223$ of a possible $224$, a normalized value of one -- a single transient state lies on the return path of nearly the entire state space, as in the collusive equilibria of Figure~\ref{fig:example-graphs}. The run-level scatter is wide, but the per-discount-factor means (large markers) trace the relationship closely -- correlating with the index at $+0.93$ for the normalized betweenness and $-1.00$ for the attractor in-degree -- so the dispersion largely reflects idiosyncratic learning noise around a monotone trend.

Note that all correlations are smaller in magnitude than on the equilibrium set. This is expected, since a Q-matrix defines a greedy action in every state, including states rarely visited during learning, so the induced graphs are noisier than equilibrium strategies. The signs and the relative ordering of the metrics nevertheless match the equilibrium analysis. Learning, moreover, does not produce the grim-trigger equilibria that Section~\ref{sec:theory} sets aside: a learned policy's collusive outcome almost always governs its dominant basin, so the collusion that emerges is of the return-to-cooperation kind the metrics are built to detect.

\section{Robustness}\label{sec:robustness}

We now test the metrics in two further settings; the details of both setups are in Appendix~\ref{app:robustness}. The first setting replaces the learning algorithm. We use the final strategy graphs of 600 runs of Decentralized Q-learning \citep{arslan2017} in the environment of Section~\ref{sec:theory} with five prices, varying the discount factor from $0.1$ to $0.9$ and holding the other parameters fixed. The algorithm learns both competitive and collusive strategy graphs, and the mean Collusion Index rises with the discount factor, from $-0.03$ at $\delta = 0.1$ to $0.56$ at $\delta = 0.9$. The correlations match the hypotheses, with a maximum betweenness of $+0.66$, an attractor in-degree of $-0.41$, and an average path length of $+0.55$; Table~\ref{tab:decq-delta} in Appendix~\ref{app:robustness} reports them by discount factor. The pattern is therefore not specific to the algorithm of Section~\ref{sec:calvano}. It also holds for an algorithm with convergence guarantees and on a coarser grid. Because Decentralized Q-learning requires increasingly large batch sizes to converge as the discount factor rises, holding the batch size fixed eventually breaks convergence at the top of the range; the batch size and the other hyperparameters should instead be varied together with the discount factor. We run this variant as well and find that our results continue to hold; see Table~\ref{tab:decq} in Appendix~\ref{app:robustness} and the description there.

The second setting keeps the algorithm and the discount factor of the collusive baseline but varies the outcome instead, using the rematching design of \citet{eschenbaum2022}. Learned policies overfit to the rival they trained with, and when two policies from distinct runs are matched against each other, collusion breaks down. We form 500 such pairs from the $\gamma = 0.95$ runs of Section~\ref{sec:calvano} and pool them with the 500 training runs. The pooled panel spans competitive and collusive outcomes at a single discount factor -- the mean index is $0.65$ among training runs and $0.14$ among rematched pairs -- and so provides an alternative to the discount-factor variation. Table~\ref{tab:three-process} shows the correlations for all four settings.

\begin{table}[h]
\centering
\begin{tabular}{lcccc}
\toprule
 & \multicolumn{4}{c}{$\rho(\cdot, \Delta)$} \\
\cmidrule(lr){2-5}
Metric & Equilibria & Calvano & DecQ & Rematch \\
\midrule
Max.\ betweenness centrality & $+0.941$ & $+0.669$ & $+0.663$ & $+0.686$ \\
Max.\ betweenness (norm.) & $+0.931$ & $+0.662$ & $+0.605$ & $+0.629$ \\
In-degree centrality & $-0.829$ & $-0.666$ & $-0.414$ & $-0.600$ \\
Number of attractors & $-0.746$ & $-0.131$ & $-0.479$ & $-0.475$ \\
Basin fraction & $+0.833$ & $+0.135$ & $+0.559$ & $+0.376$ \\
Closeness centrality & $-0.659$ & $-0.334$ & $-0.004$ & $+0.324$ \\
Average path length & $+0.869$ & $+0.566$ & $+0.547$ & $+0.183$ \\
\midrule
$\bar\Delta$ & $0.580$ & $0.385$ & $0.336$ & $0.396$ \\
$N$ & 101 & 1800 & 600 & 1000 \\
\bottomrule
\end{tabular}
\caption{Pearson correlations $\rho(\cdot,\Delta)$ between strategy graph metrics and the Collusion Index: the analytically known Nash equilibria of the three-price game (\emph{Equilibria}), Calvano Q-learning with fifteen prices pooled over discount factors $\gamma\in\{.1,.3,.5,.7,.9,.95\}$, Decentralized Q-learning with five prices, and a panel pooling the 500 training runs at $\gamma=0.95$ with 500 rematched pairs of independently trained policies (\emph{Rematch}).}
\label{tab:three-process}
\end{table}

Two results stand out. First, the maximum betweenness and the attractor in-degree carry the hypothesized signs in all four settings, at magnitudes between $0.41$ and $0.94$; the average path length does as well, though weakly in the rematch panel ($+0.18$). Second, the basin fraction and the number of attractors, uninformative in the Calvano case, are informative for Decentralized Q-learning ($+0.56$ and $-0.48$) and in the rematch panel ($+0.38$ and $-0.48$). This is the pattern the equilibrium analysis predicts. The two metrics separate the classes only through fragmented competitive graphs (Section~\ref{sec:theory}), and the settings differ in exactly this respect: in the Calvano case competitive policies learned at low discount factors converge to single-basin graphs like panel~(a) of Figure~\ref{fig:example-graphs} ($1.1$ attractors on average), while the rematched pairs combine policies that do not fit one another and fragment ($2.4$ attractors on average, against $1.3$ among training runs), as do the low-discount Decentralized Q-learning runs. These two metrics thus track the fragmentation of the state space, whatever its source.

The closeness centrality is the one metric that is not consistent across settings: negative on the equilibrium set and under the discount-factor sweep, essentially zero for Decentralized Q-learning, but positive in the rematch panel ($+0.32$). Within the rematched pairs alone, where nearly all outcomes are competitive, only the metrics defined on the attractor itself remain correlated with the index (Appendix~\ref{app:robustness}).

\section{Discussion and Policy}\label{sec:discussion}

The policy implication concerns the scope of access needed for an audit. An authority need not reconstruct how an algorithm was trained if it can query a fixed snapshot of the deployed policy. The authority would identify the policy's state variables, feasible states, actions, memory, and state-transition rule, then obtain the action selected by each policy for every feasible joint state in $\hat{\mathcal{S}}$. The resulting successor map defines the strategy graph. Maximum betweenness and attractor in-degree would likely provide the primary screen. Because these calculations use only unlabeled topology, a response interface or sandbox could require significantly less disclosure than source code, training data, and internal model parameters. However, the authority would still need to verify that the queried policy is the one used in the market.

This screen is not a finding of collusion. The analysis does not provide a market-independent cutoff, and a bottleneck may reflect an operational constraint or an arbitrary response in a state rarely encountered during training. An unlabeled graph also does not establish that an attractor has a high price, nor can it identify agreement, intent, or competitive harm. Its natural role is to complement outcome-based evidence. Prices and quantities describe realized conduct and market effects; the policy graph describes how the algorithms would respond away from the observed path. A structural signal can therefore direct the authority toward particular states and policy responses for closer economic and institutional analysis.

Practical use also requires moving beyond finite, deterministic, and frozen policies with a unique greedy action. A deployed policy may continue to update, randomize, use continuous actions, or condition on variables unavailable to the auditor. A stochastic policy induces a transition matrix rather than a functional graph, while a snapshot of an adaptive policy omits its subsequent learning. The number of queries can also become prohibitive. If each of $n$ firms chooses from $m$ prices and conditions on $l$ past price profiles, the joint history contains $m^{nl}$ states. Sampling or compressing the state space would reduce this burden, but may miss precisely the off-path transitions that reveal a bottleneck. On the other hand, such state spaces are also prohibitively large from the perspective of the learning algorithms themselves. It is thus unlikely that they will use the memory state space, but would likely make use of lower dimensional representations.

\section{Conclusion}\label{sec:conclusion}

This paper asks whether collusive structure can be recognized from pricing policies without knowing demand or competitive and monopoly benchmarks. We derive five graph-theoretic metrics from the complete, analytically known equilibrium set of a three-price game, where the strategy graph of a reward-and-punishment strategy can be compared directly with that of competition, and take these metrics as hypotheses to policies produced by three learning procedures. Across these settings the maximum betweenness and the attractor in-degree provide consistent signals of supra-competitive outcomes, and the average path length is also informative in most of them. The intuition is simple. Collusive policies concentrate their return paths through a small number of disciplining states and give fewer states direct access to the attractor. This is a topological signature of reward and punishment: it is visible in the unlabeled transition graph alone, without the prices, profits, or benchmarks that define the Collusion Index against which we validate it.

For an auditor, analyzing the strategy graph requires access to the trained algorithm, but not to its codebase or training data which may be infeasible or hard to interpret. The benefit is the ability to assess responses off a realized price path. Having access only to a sequence of actions (such as the posted prices) implies observing a single trajectory through a policy but not how it would respond off that path. Querying a frozen policy and reading its graph in turn requires neither observed prices nor the underlying codebase, and the metrics use only the topology of greedy responses. Their natural role is a benchmark-free structural screen that can complement outcome-based evidence.

Two limitations of our analysis should be considered. First, the metrics detect collusion that returns to cooperation after punishment. Hence, an equilibrium that punishes by reverting permanently to competition -- a grim trigger -- hides its collusive branch in a subordinate component and is indistinguishable from competition on the largest weakly connected component. Almost all competitive-class equilibria of the three-price game are of this kind, even if learned policies rarely are. Second, the validation is so far confined to finite, deterministic, frozen policies in stylized logit environments. Stochastic, adaptive, or high-dimensional policies, and markets with more than two firms, remain open. Extending the metrics to these settings, calibrating them into thresholds with known statistical properties, and validating them against deployed pricing systems are the natural next steps.

\bibliographystyle{apalike}
\bibliography{references}

\newpage
\appendix

\section{Robustness Test Details}\label{app:robustness}

\paragraph{Attractor convention and collusion index} The metric validation of Section~\ref{sec:theory} rests on two modelling choices, and Table~\ref{tab:sensitivity} checks both against the convention used in the paper, on the $101$ equilibria. The first is which attractor to read the markers off when a strategy graph has more than one. We use the largest weakly connected component, the outcome that governs play from the dominant basin; the alternative is the highest-price attractor, the collusive outcome where one exists. Under the largest-WCC convention every marker carries its hypothesized sign, whereas under the highest-price convention the maximum betweenness ($+0.02$), the average path length ($-0.03$) and the basin fraction ($-0.25$) lose their signal and the number of attractors changes sign, leaving only the in-degree ($-0.69$) and closeness ($-0.51$). The reason is the grim-trigger structure of Section~\ref{sec:theory}: for those equilibria the highest-price attractor is a subordinate component, an isolated collusive state into which no punishment path leads, so the reward-punishment structure the markers detect lies in the dominant competitive basin rather than around the collusive outcome itself. The second choice is the Collusion Index. Throughout we use the profit-based index; replacing it with a price-based one -- the average price at the attractor normalized by the Nash and monopoly prices, linear rather than concave in the price level -- leaves the correlations essentially unchanged (third column of Table~\ref{tab:sensitivity}). The markers thus depend on reading structure off the dominant basin, but not on how the collusive outcome is scored.

\begin{table}[h]
\centering
\begin{tabular}{lccc}
\toprule
 & \multicolumn{3}{c}{$\rho(\cdot, \Delta)$} \\
\cmidrule(lr){2-4}
Metric & Largest WCC & Highest price & Price-based $\Delta$ \\
\midrule
Max.\ betweenness centrality & $+0.941$ & $+0.025$ & $+0.919$ \\
In-degree centrality & $-0.829$ & $-0.693$ & $-0.839$ \\
Number of attractors & $-0.746$ & $+0.206$ & $-0.714$ \\
Basin fraction & $+0.833$ & $-0.249$ & $+0.764$ \\
Closeness centrality & $-0.659$ & $-0.506$ & $-0.709$ \\
Average path length & $+0.869$ & $-0.031$ & $+0.907$ \\
\midrule
$\bar\Delta$ & $0.580$ & $0.846$ & $0.485$ \\
$N$ & 101 & 101 & 101 \\
\bottomrule
\end{tabular}
\caption{Pearson correlations $\rho(\cdot,\Delta)$ on the 101 Nash equilibria of the three-price game, testing the two modelling choices behind the validation. The first column is the convention used throughout the paper (markers on the largest weakly connected component, profit-based Collusion Index); the second selects the highest-price attractor instead of the largest WCC; the third replaces the profit-based index with a price-based one (the average price normalized by the Nash and monopoly prices).}
\label{tab:sensitivity}
\end{table}

\paragraph{Price grid and discount factor} Two further variations of the Calvano Q-learning environment of Section~\ref{sec:calvano} leave the pattern intact; Table~\ref{tab:calvano-robustness} reports both, computed with the same markers and the largest-WCC convention. The first pools the $500$ baseline runs at $\gamma = 0.95$ with the $500$ low-discount runs at $\gamma = 0.1$ into a single competitive-versus-collusive panel ($N = 1000$); this two-point contrast gives the strongest correlations for the two leading markers, a maximum betweenness of $+0.76$ and an attractor in-degree of $-0.75$. The second widens the price grid to twenty-one prices and lengthens training to $T = 2 \times 10^6$ periods, again at $\gamma = 0.95$ ($N = 500$). The wider grid raises the mean Collusion Index from $0.65$ to $0.86$, and the hypothesized signs survive, though at weaker magnitudes, because almost every run is now collusive and the index varies little across them.

\begin{table}[h]
\centering
\begin{tabular}{lcc}
\toprule
 & \multicolumn{2}{c}{$\rho(\cdot, \Delta)$} \\
\cmidrule(lr){2-3}
Metric & Baseline\,+\,low-$\gamma$ & Wider grid \\
\midrule
Max.\ betweenness centrality & $+0.760$ & $+0.596$ \\
Max.\ betweenness (norm.) & $+0.749$ & $+0.586$ \\
In-degree centrality & $-0.750$ & $-0.350$ \\
Number of attractors & $-0.121$ & $-0.194$ \\
Basin fraction & $+0.142$ & $+0.100$ \\
Closeness centrality & $-0.306$ & $+0.192$ \\
Average path length & $+0.588$ & $+0.272$ \\
\midrule
$\bar\Delta$ & $0.425$ & $0.859$ \\
$N$ & 1000 & 500 \\
\bottomrule
\end{tabular}
\caption{Pearson correlations $\rho(\cdot,\Delta)$ for two further variations of the Calvano Q-learning environment: the baseline discount factor $\gamma=0.95$ pooled with the low-discount runs $\gamma=0.1$, and a wider grid of twenty-one prices trained for $2\times10^6$ periods at $\gamma=0.95$. Markers computed on the largest-WCC attractor.}
\label{tab:calvano-robustness}
\end{table}

\paragraph{Decentralized Q-learning} The algorithm is the Decentralized Q-learning of \citet{arslan2017}, analyzed for pricing games in \citet{meylahn2025}. The environment is the logit duopoly of Section~\ref{sec:theory}. The five prices are equally spaced on the price grid of \citet{calvano2020} with $\xi = 0.1$, which contains neither $p^N$ nor $p^M$ exactly; the two grid prices closest to $p^N$ correspond to index values of $-0.16$ and $+0.32$. Firms have a memory of one period and observe both prices, so the joint state space consists of the $25$ price pairs. We observe the final strategy graph of each run rather than play, and compute the Collusion Index from the profits at the attractor states. In the primary specification the batch size ($10^6$, over $300$ batches) and the exploration rate ($0.15$) are held fixed and only the discount factor is varied, over $\{0.1, 0.3, 0.5, 0.7, 0.8, 0.9\}$ with $100$ runs each; the mean index rises from $-0.03$ at $\delta = 0.1$ to $0.56$ at $\delta = 0.9$ (Table~\ref{tab:decq-delta}). We omit $\delta = 0.95$, at which every run collapses to the competitive sink under this batch size. As a further check we also analyze $500$ runs in which the exploration rate, batch size, and discount factor vary jointly, giving an index between $-0.16$ and $0.95$ with mean $0.41$ and unchanged marker signs (Table~\ref{tab:decq}).

\begin{table}[h]
\centering
\scriptsize
\begin{tabular}{lccccccc}
\toprule
 & \multicolumn{7}{c}{$\rho(\cdot, \Delta)$} \\
\cmidrule(lr){2-8}
Metric & $\delta{=}0.1$ & $\delta{=}0.3$ & $\delta{=}0.5$ & $\delta{=}0.7$ & $\delta{=}0.8$ & $\delta{=}0.9$ & Pooled \\
\midrule
Max.\ betweenness centrality & $+0.613$ & -- & $+0.581$ & $+0.728$ & $+0.770$ & $+0.647$ & $+0.663$ \\
Max.\ betweenness (norm.) & $+0.529$ & -- & $+0.586$ & $+0.650$ & $+0.721$ & $+0.626$ & $+0.605$ \\
In-degree centrality & $-0.436$ & -- & $-0.342$ & $-0.554$ & $-0.613$ & $-0.485$ & $-0.414$ \\
Number of attractors & $-0.736$ & -- & $-0.070$ & $-0.319$ & $-0.403$ & $-0.231$ & $-0.479$ \\
Basin fraction & $+0.463$ & -- & $+0.013$ & $+0.266$ & $+0.371$ & $+0.324$ & $+0.559$ \\
Closeness centrality & $+0.125$ & -- & $-0.170$ & $-0.171$ & $-0.207$ & $+0.139$ & $-0.004$ \\
Average path length & $+0.449$ & -- & $+0.426$ & $+0.565$ & $+0.656$ & $+0.404$ & $+0.547$ \\
\midrule
$\bar\Delta$ & $-0.032$ & $0.323$ & $0.325$ & $0.407$ & $0.435$ & $0.556$ & $0.336$ \\
$N$ & 100 & 100 & 100 & 100 & 100 & 100 & 600 \\
\bottomrule
\end{tabular}
\caption{Pearson correlations $\rho(\cdot,\Delta)$ between strategy graph metrics and the Collusion Index for Decentralized Q-learning by discount factor $\delta$ (five-price Calvano logit setting; batch size, exploration rate and memory held fixed). Markers computed on the largest-WCC attractor.}
\label{tab:decq-delta}
\end{table}

\begin{table}[h]
\centering
\begin{tabular}{lc}
\toprule
 & \multicolumn{1}{c}{$\rho(\cdot, \Delta)$} \\
\cmidrule(lr){2-2}
Metric & DecQ (5 prices) \\
\midrule
Max.\ betweenness centrality & $+0.761$ \\
In-degree centrality & $-0.638$ \\
Number of attractors & $-0.063$ \\
Basin fraction & $-0.013$ \\
Closeness centrality & $-0.618$ \\
Average path length & $+0.775$ \\
\midrule
$\bar\Delta$ & $0.409$ \\
$N$ & 500 \\
\bottomrule
\end{tabular}
\caption{Pearson correlations $\rho(\cdot,\Delta)$ between strategy graph metrics and the Collusion Index for the Decentralized Q-learning runs (5-price Calvano logit setting; exploration rate, batch size and discount factor varied across runs). Markers computed on the largest-WCC attractor.}
\label{tab:decq}
\end{table}

\paragraph{Rematched policies} For each of 500 pairs $(i, j)$ of distinct $\gamma = 0.95$ runs from Section~\ref{sec:calvano}, we match the first agent's Q-matrix from run $i$ with the second agent's Q-matrix from run $j$. The pair plays $1{,}000$ periods from a random initial state, without exploration and without updating. The Collusion Index is computed from the final 100 periods and the metrics from the joint strategy graph of the two Q-matrices. The mean index across the rematched pairs is $0.14$, with a standard deviation of $0.14$. Within the rematched pairs alone, the attractor in-degree ($-0.46$) and the closeness centrality ($-0.38$) remain correlated with the index, the average path length only weakly ($+0.21$), and the maximum betweenness not at all ($+0.06$): rematched graphs retain sizable bottlenecks from the two collusive policies they combine -- the mean maximum betweenness is $89$ of a possible $224$ -- even where realized play is competitive. The rematch panel of Table~\ref{tab:three-process} pools these 500 pairs with the 500 training runs.

\section{Equilibrium-Set Correlations}\label{app:additionalTables}

Table~\ref{tab:theory} splits the equilibrium-set correlations of Section~\ref{sec:theory} into the 29 symmetric and 72 asymmetric equilibria. Every marker keeps its hypothesized sign in both subsets and the magnitudes are close, so the combined correlations reported in the main text are not an artifact of pooling the two.

\begin{table}[h]
\centering
\begin{tabular}{lccc}
\toprule
 & \multicolumn{3}{c}{$\rho(\cdot, \Delta)$} \\
\cmidrule(lr){2-4}
Metric & Symmetric & Asymmetric & Combined \\
\midrule
Max.\ betweenness centrality & $+0.912$ & $+0.951$ & $+0.941$ \\
In-degree centrality & $-0.756$ & $-0.856$ & $-0.829$ \\
Number of attractors & $-0.711$ & $-0.756$ & $-0.746$ \\
Basin fraction & $+0.755$ & $+0.871$ & $+0.833$ \\
Closeness centrality & $-0.535$ & $-0.697$ & $-0.659$ \\
Average path length & $+0.866$ & $+0.874$ & $+0.869$ \\
\midrule
$\bar\Delta$ & $0.462$ & $0.628$ & $0.580$ \\
$N$ & 29 & 72 & 101 \\
\bottomrule
\end{tabular}
\caption{Pearson correlations $\rho(\cdot,\Delta)$ between strategy graph metrics and the Collusion Index on the analytically known Nash equilibria of the three-price game (Calvano logit baseline, $\xi=0$). Metrics computed on the largest-WCC attractor; $\Delta$ profit-based.}
\label{tab:theory}
\end{table}

\end{document}